\documentclass[12pt,preprint]{aastex}
\def\beq{\begin{equation}}
\def\eeq{\end{equation}}
\def\ben{\begin{eqnarray}}
\def\een{\end{eqnarray}}

\def\jd{{\bf j}_{d}}
\def\jl{{\bf j}_{cg}}
\def\jh{{\bf j}_{hg}}
\def\js{{\bf j}_{s}}
\def\us{{\bf u}_{s}}
\def\ul{{\bf u}_{cg}}
\def\uh{{\bf u}_{hg}}
\def\ud{{\bf u}_{d}}
\def\em{{\bf e}_{1}}
\def\ei{{\bf e}_{2}}
\def\en{{\bf e}_{3}}
\def\tris{{\cal T}_{s}}
\def\tril{{\cal T}_{cg}}
\def\trih{{\cal T}_{hg}}
\def\trid{{\cal T}_{d}}
\def\sphs{{\cal S}_{s}}
\def\sphl{{\cal S}_{cg}}
\def\sphh{{\cal S}_{hg}}
\def\sphd{{\cal S}_{d}}

\def\am{a(z_{m})}

\begin{document}
\title{Merger Effects on the Spin and Shape Alignments of Galaxy Stellar, Cold and Hot Gas, and Dark Matter Components}
\author{Jounghun Lee and Jun-Sung Moon}
\affil{Astronomy program, Department of Physics and Astronomy,
Seoul National University, Seoul  08826, Republic of Korea \\
\email{jounghun@astro.snu.ac.kr}}

\begin{abstract}
We present a numerical evidence supporting the scenario that the peculiar alignments of the galaxy stellar spins with the major principal 
axes of the local tidal tensors are produced during the quiescent evolution period when the galaxies experience no recent merger events. 
Analyzing the merger tree from the TNG300-1 simulation of the IllustrisTNG project, we find the latest merger epochs, 
$\am$, of the galaxies, and create four $\am$-selected samples that are controlled to share the identical mass and density distributions. 
For each sample, we determine the spin and shape vectors of the galaxy stellar, cold and hot gas, and dark matter components separately, 
and compute the average strengths of their alignments with the principal directions of the local tidal fields as well as their mutual alignment tendencies. 
It is found that the stellar (cold gas) spin axes of the galaxies whose latest merger events occur at earlier epochs are more strongly 
aligned (weakly anti-aligned) with the major principal axes of the tidal fields. 
It is also shown that although the mass-dependent transition of the galaxy DM spins have little connection with the merger events, 
the morphologies, spin-shape and shape-shear alignment strengths of the galaxy four components sensitively depend on $\am$. 
Noting that the stellar components of the galaxies which undergo long quiescent evolution have distinctively oblate shapes and very strong 
spin-shape alignments, we suggest that the local tidal field might be traced by using the stellar shapes of galaxies without signatures of mergers
as a proxy of their stellar spins. 
\end{abstract}
\section{Introduction}\label{sec:intro}

The {\it peculiar tidal connection of the galaxy stellar spins} is a term first coined by Lee, Moon \& Yoon (2022) to describe the phenomenon 
detected in a hydrodynamical simulation that the spin axes of the galaxy stellar components tend to be aligned with the major 
principal directions of the local tidal fields. This phenomenon is peculiar since the non-stellar gas and dark matter (DM) components of the galaxies 
strongly prefer the directions {\it perpendicular} to the major principal axes of the local tidal tensors in their spin orientations, regardless of redshifts 
and mass scales \citep{lee-etal21}.  Although \citet{lee-etal22} showed that this peculiar phenomenon must be closely linked with a physical process 
responsible for the deviation of the galaxy stellar spin directions from their DM counterparts, its origin is still shrouded in mystery, calling for more 
probing attentions and follow-up works.

The detection of this peculiar phenomenon has casted a doubt on the conventional assumption that although the spin axes of the galaxy stellar 
components are not perfectly in parallel with those of the DM counterparts \citep[e.g.,][]{hah-etal10,vel-etal15,ten-etal17,zju-etal17,sou-etal20}, 
non-random orientations of the former in the cosmic web must reflect the intrinsic alignments of the latter \citep[e.g.,][and references therein]{cod-etal18}, 
the existence of which have been theoretically predicted and numerically confirmed  \citep[for a comprehensive review, see][]{align_review15}. 
In fact, this conventional assumption was adopted by many observational endeavors to find evidences for the effect of the cosmic web on the evolution 
of the angular momentum of the galactic halos by measuring the galaxy stellar spin orientations relative to the cosmic web.

For example, \citet{nav-etal04} showed that the observed spin axes of the spiral galaxies embedded in the Local Superclusters (LS) are inclined toward the 
directions parallel to the LS plane, and suggested their observations as an evidence supporting the linear tidal torque theory which generically predicts the 
alignments between the spin axes of the galactic halos and the intermediate principal axes of the tidal tensors \citep[see also][]{tru-etal06,LE07}.
\citet{jon-etal10} found a tendency of the spin vectors of the edge-on spiral galaxies being aligned with the directions perpendicular to the filaments 
in the local universe  and interpreted this as a "fossil evidence" of the existence of the large-scale tidal effect on the angular momentum vectors of the 
galactic halos \citep[see, also][]{kra-etal21}.
\citet{mot-etal21} measured a significant signal of the alignments between the spin axes of the local spirals and the principal axes of the initial tidal 
field reconstructed from the spatial distributions of the local galaxies from an all-sky survey, and declared it as a detection of the effect of the initial tidal torques 
on the angular momentum of proto-galactic halos \citep{whi84,LP00,LP01}.

However, if the observed non-random orientations of the galaxy stellar spins with respect to the cosmic web do not take on the essential aspect of the 
theoretically predicted intrinsic spin alignments of the galactic halos, as hinted by \citet{lee-etal22}, then it requires a modification of the conventional assumption 
and a new interpretation of the observational evidences, which of course, should be preceded by physically understanding how the galaxy stellar spins acquire the 
peculiar tidal connection. Among many possible factors that may affect the evolution of the galaxy stellar spin vectors and their deviations from the DM counterparts, 
here we consider the merging events that most of the galaxies experience in their growths, in light of the previous results based on hydrodynamical simulations 
that the alignment strength and tendency of the galaxies depend strongly on their dynamics and progenitor histories \citep{chi-etal16,bho-etal20,jag-etal22}. 

If the merging events make a substantial contribution to or at least have a link with the generation of the peculiar spin alignments of the galaxy stellar components, then the 
alignment strengths should sensitively depend on the latest merger epochs.  
In this Paper, we are going to explore if and how the peculiar stellar  spin and shape alignments of the galaxies at $z=0$ vary with their latest merger epochs with the 
help of a high-resolution hydrodynamical simulation. Here, we focus only on the galaxies at the present epoch, $z=0$, since the signal of the peculiar stellar spin alignments 
was found to be significant only at low redshifts, $z<0.5$ \citep{lee-etal22}. In addition, we will also inspect the effects of merger events on the spin and shape alignments 
of the galaxy DM, cold and hot gas components for comparison. 

\section{Numerical data and analysis}\label{sec:data}

We utilize the data from the TNG300-1 simulation belonging to a series of magnetohydrodynamical IllustrisTNG project
\citep{tngintro1, tngintro2, tngintro3, tngintro4, tngintro5, illustris19} conducted for a $\Lambda$CDM cosmology 
with the Planck initial conditions \citep{planck16}. Employing the elaborate code, Arepo \citep{arepo}, the IllustrisTNG simulations were 
capable of incorporating the full non-gravitational baryonic processes known to play the key roles in driving the galaxy evolution. 
A detailed description of the IllustrisTNG project can be found in its web page\footnote{https://www.tng-project.org/data/}. 

The TNG300-1 simulation was performed in a periodic box of linear  size $302.6\,$Mpc containing $2500^{3}$ baryonic gas and equal number of DM particles, 
the mass resolution of which are as high as $1.1\times 10^{7}\,M_{\odot}$ and $5.9\times 10^{7}\,M_{\odot}$, respectively. 
It found the substructures of the friends-of-friends (FoF) groups via the Subfind algorithm \citep{subfind} and constructed their merger trees via the SubLink code \citep{sublink},  
the catalogs of which can be extracted from the aforementioned webpage. Hereafter, we will refer to the substructures of the FoF groups as the galaxies. 

With the help of the routine provided by \citet{lee-etal21}, we first construct the Gaussian-filtered tidal field, $T_{ij}({\bf x})$, on the $256^{3}$ grid points at $z=0$.
A brief summary of the routine is as follows: (i) Constructing the mass density contrast field, $\delta({\bf x})$, on the $256^{3}$ grid points by 
applying the cloud-in-cell method to the FoF halo sample with no mass-cut at $z=0$. (ii) Performing the Fourier transformation of $\delta({\bf x})$ 
to obtain its Fourier amplitude $\delta({\bf k})$ by using the Fast Fourier Transformation (FFT) code \citep{pre-etal92}. (iii) Computing the Fourier 
amplitude of the Gaussian-filtered tidal field as $T_{ij}({\bf k})=\left[k_{i}k_{j}\delta({\bf k})/k^{2}\right]\exp(-k^{2}R^{2}_{f}/2)$ where 
$k=\vert{\bf k}\vert$ and $R_{f}$ is the filtering radius. (iv) Performing the inverse Fourier transformation of $T_{ij}({\bf k})$ to obtain $T_{ij}({\bf x})$. 
As in \citet{lee-etal22}, we set the filtering radius at $R_{f}=1.5\,$Mpc,  the median size of cluster environments in which the majority 
of the galaxies are embedded  \citep[e.g.,][]{sch-etal97,tem-etal14} .
 At the grid that matches the position of each galaxy, we carry out the similarity transformation of $T_{ij}({\bf x})$ to find its major, intermediate 
and minor principal axes ($\em,\ \ei$ and $\en$, respectively), corresponding to the largest, second largest and smallest eigenvalues 
($\lambda_{1},\ \lambda_{2}$ and $\lambda_{3}$, respectively). 
The major (minor) principal axes of the local tidal tensors are parallel to the directions of maximum (minimum) matter compression. 

For the investigation of the merger effects on the spin and shape alignments of the galaxies with the principal directions of the local tidal fields, we separately 
consider the four components of each galaxy, namely, stellar, cold gas, hot gas and DM.  Here, the cold (hot) gas component corresponds to the non-stellar baryons with 
temperature, $T$, lower (higher) than $10^{5}$ K, above which the gas is mostly created by gravitational shock heating or feedback processes \citep[e.g.,][]{mar-etal19}.  
We determine the temperature of each non-stellar gas cell by converting its specific internal energy and electron abundance, information on which are all available in the 
TNG300-1 snapshot data. 
The public data release from the IllustrisTNG project also provides all other necessary information for our investigation, such as the comoving positions and velocities of 
all member particles, numbers of the cold gas, hot gas and DM components of each galaxy ($n_{cg}$, $n_{hg}$, and $n_{d}$, respectively), and total mass of all member 
particles ($M_{\rm tot}$), and numbers of the stellar cells ($n_{s}$) within twice its {\it stellar} half-mass radius  ($2R_{1/2}$) at $z=0$. 

For each of the galaxies with $n_{s}\ge 300$, we compute the spin vector of its stellar component, ${\bf J}_{s}$, as
\begin{equation}
\label{eqn:star_spin}
{\bf J}_{s} = \sum_{\alpha=1}^{n_{s}}{m}_{s,\alpha}\,[({\bf x}_{s,\alpha}-{\bf x}_{c})\times({\bf v}_{s,\alpha}-{\bf v}_{c})]\, ,\\
\end{equation}
where $m_{s,i}$, ${\bf x}_{s,\alpha} = (x_{s,\alpha\,i})$ and ${\bf v}_{s,\alpha}$ are the mass, comoving position and peculiar velocity of the $\alpha$th 
stellar particle within $2R_{1/2}$, while ${\bf x}_{c}=(x_{c,i})$ and ${\bf v}_{c}$ are the positions and velocities of the galaxy center, respectively. 
The number cut of the stellar particles, $n_{s}\ge 300$, is applied to the galaxies, since the measurements of the stellar spin vectors from the lower number 
of particles would suffer from low accuracy \citep{bet-etal07}. The unit stellar spin vector, $\js$,  is then calculated as $\js\equiv {\bf J}/\vert{\bf J}\vert$. 

The stellar inertia tensor, ${\bf I}_{s}=(I_{s,ij})$, of each selected galaxy with $n_{s}\ge 300$ is computed as \citep{CL96}: 
\begin{equation}
\label{eqn:star_iner}
I_{s,ij} = \frac{1}{M_{s}}\sum_{\alpha=1}^{n_{s}}{m}_{s,\alpha}\,(x_{s,\alpha\,i}-x_{c,i})\, (x_{s,\alpha\,j}-x_{c,j})\, , \quad  i,j \in \{1,\ 2,\ 3\}\, ,
\end{equation}
where $M_{s}\equiv \sum_{\alpha=1}^{n_{s}}{m}_{s,\alpha}$. 
The similarity transformation of ${\bf I}_{s}$ yields a set of orthonormal eigenvectors corresponding to its three eigenvalues (say, $q_{1},q_{2},q_{3}$ in a decreasing order).
We define the unit stellar shape vector of a galaxy, $\us$, as the eigenvector of ${\bf I}_{s}$ corresponding to $q_{3}$, which is in parallel to the direction of the minor axis 
of the galaxy stellar shape. We also quantify the morphological shape of the stellar distribution of each selected galaxy by measuring its degrees of triaxiality ($\tris$) and 
sphericity ($\sphs$) \citep{fra-etal91,bet-etal07}:
\begin{equation}
\label{eqn:star_tris}
{\cal T}_{s} = \frac{q_{1}-q_{2}}{q_{1}-q_{3}},\qquad {\cal S}_{s} = \left(\frac{q_{3}}{q_{1}}\right)^{1/2}\, .
\end{equation}
The oblate (prolate) shape corresponds to $\tris< 0.5$ ($\tris> 0.5$), while the more aspherical shape has a lower value of $\sphs$.
Separately applying three different particle number cuts, $n_{cg}\ge 300$, $n_{hg}\ge 300$, $n_{d}\ge 300$, we determine the unit spin and shape vectors, 
$\{\jl,\ul\}$, $\{\jh,\uh\}$, $\{\jd,\ud\}$,  and morphological shapes, $\{\tril,\sphl\}$, $\{\trih,\sphh\}$, $\{\trid,\sphd\}$, of each selected galaxy in a similar manner. 

At this stage, it is worth explaining here why we include only those stellar components within $2R_{1/2}$, rather than using all of them contained in the host subhalos 
for the measurements of $\js$ and $I_{s,ij}$. The main reason is to make our numerical analysis in line with the observational ones. 
For the determination of the spin orientations of the real galaxies from observations, it is necessary to measure their kinetic properties. 
But, it has been known for long that measuring them outside $2R_{1/2}$ is very difficult in practice \citep{rom-etal03,coc-etal09,wel-etal20}. 
For this very reason, most of the spectroscopic galaxy surveys like the Institute For Astronomy Survey (IFA), the Mapping Nearby Galaxies at APO survey 
(MaNGA), the Sydney-AAO Multi-object Integral Field Spectrograph Galaxy Survey (SAMI), and the Calar Alto Legacy Integral Field Area Survey (CALIFA) 
were originally designed to observe the stellar components within $2R_{1/2}$ \citep{kro-etal19,lee-etal19,wel-etal20}. 
Moreover, given that the spin parameters, $\lambda$, of the real galaxies from the observations are usually measured from the stellar components within 
$2R_{1/2}$ \citep{ems-etal07,cor-etal16}, it should be quite reasonable and appropriate to adopt the same radial distance cutoff, $2R_{1/2}$, 
for the determination of the stellar spin directions \footnote{In real observations, $R_{1/2}$ is defined as the half luminosity radius rather than the half stellar mass radius 
\citep{rom-etal03}.}. 

Tracking back the main progenitor branch of the merger tree for each galaxy in the TNG300-1, we determine the scale factor of the epoch, say $a(z_{m})$,  
when its latest merger event occurs.  For this analysis, we consider only those mergers events in which the involved galaxies have stellar mass ratios 
larger than $0.1$ (i.e., both major and minor mergers) and exclude those galaxies which never experience any merger event during the evolution. 
The merger mass ratio is calculated from the past maximum stellar masses of each merging galaxy to avoid the notorious artificial mass loss problem in the 
halo finding algorithm \citep{sublink}. Splitting the galaxies into four equal-size samples according to the values of $\am$,  
we control the four samples to have identical joint distributions of $\log\tilde{M}_{\rm tot}$ and $\log\,(1+\delta)$, where 
$\tilde{M}_{\rm tot}\equiv M_{\rm tot}/(10^{7}\,h^{-1}\,M_{\odot})$ and $\delta\equiv \sum_{i=1}^{3}\lambda_{i}$, for the purpose of nullifying the possible effects 
of the differences in the total masses of the galaxies and the environmental densities among the four samples on the alignment strengths, as done in \citet{lee-etal22}. 

We calculate the following alignments ensemble averaged over each of the controlled samples (say, $A$, $B$, $C$ and $D$),
and investigate the variations of their strengths with $\am$: the spin alignments between the baryon and DM components
($\langle\js\cdot\jd\rangle$, $\langle\jl\cdot\jd\rangle$, $\langle\jh\cdot\jd\rangle$); 
the alignments between the baryon spins and major principal axes of the local tidal tensors called {\it the spin-shear alignments} 
($\langle\vert\js\cdot\em\vert\rangle$, $\langle\vert\jl\cdot\em\vert\rangle$, $\langle\vert\jh\cdot\em\vert\rangle$); 
and the alignments between the DM spins of high-mass (low-mass) galaxies and intermediate (minor) principal axes of the local tidal tensors 
($\langle\vert\jd\cdot\em\vert\rangle$, $\langle\vert\jd\cdot\ei\vert\rangle$, $\langle\vert\jd\cdot\en\vert\rangle$); 
the alignments between the spin and shape vectors of each component called {\it the spin-shape alignments
($\langle\vert\js\cdot\us\vert\rangle$, $\langle\vert\jl\cdot\ul\vert\rangle$, $\langle\vert\jh\cdot\uh\vert\rangle$, $\langle\vert\jd\cdot\ud\vert\rangle$);}
the alignments between the shape vectors and tidal major principal directions called {\it the shape-shear alignments}
($\langle\vert\us\cdot\em\vert\rangle$, $\langle\vert\ul\cdot\em\vert\rangle$, $\langle\vert\uh\cdot\em\vert\rangle$, $\langle\vert\ud\cdot\em\vert\rangle$). 
We also compute the standard deviation in each ensemble average as the associated error. For instance, the error in $\langle\js\cdot\jd\rangle$ is computed as 
$\sqrt{\left[\langle(\js\cdot\jd)^{2}\rangle - \langle\js\cdot\jd\rangle^{2}\right]/(N_{g}-1)}$ where $N_{g}$ is the number of the galaxies belonging to each sample.

Note that the spin alignments between the baryon and DM components are measured as the dot-products between the spin vectors while the spin-shear alignments are 
measured as the {\it absolute values} of the dot-products between the galaxy spin vectors and the principal axes of the local tidal tensors. 
It is simply because for the latter case there is no difference in the physical meaning between the positive and negative signs of the dot-product, while for the former case 
there is.  
For instance, suppose that there is a galaxy whose stellar components are spinning clockwise when viewed face on. If $\js\cdot\jd=-1$, then it indicates counter-clockwise 
spinning of the galaxy DM components. It is also worth mentioning here that the criteria on $a_{m}(z)$ used to separate the four samples depend on which quantity to 
calculate, since different numbers of galaxies are selected by different particle number cuts to calculate the alignments of different components. 
For example, for the calculation of $\langle\js\cdot\jd\rangle$, only those galaxies that satisfy both of  the criteria of $n_{s}\ge 300$ and $n_{d}\ge 300$ are selected and 
divided into the four samples. For the calculation of $\langle\vert\js\cdot\em\vert\rangle$, the single condition of $n_{s}\ge 300$ is used to separate the samples.

\section{Dependence on the latest merger epochs}\label{sec:bdm}

\subsection{Baryon spin-DM spin alignments} 

Figure \ref{fig:jjd} shows how the mean strengths of the alignments of the stellar, cold and hot gas spin axes of the galaxies with their DM counterparts  
depend on the latest merger epochs, $\am$, at $z=0$ (top, middle and bottom panels, respectively). As can be seen, the earlier the latest merger events 
occur, the lower the mean alignments between the galaxy baryon and DM spin vectors become. Note that for the case of the galaxies with $\am<0.4$, 
the value of $\langle\js\cdot\jd\rangle$ drops below $0.5$, the expectation value from two randomly oriented unit vectors. 
From here on, we refer to this tendency of having the ensemble average of the dot-product between two unit vectors below $0.5$ as {\it anti-alignment}. 
The result shown in the top panel of Figure \ref{fig:jjd} implies that the mechanism responsible for this $\js$-$\jd$ anti-alignment must occur only in the 
quiescent period of time when no merging event occurs. 

Unlike the stellar components, both of the cold and hot gas components exhibit no anti-alignment but only alignment tendency with the DM counterparts 
($\langle\jl\cdot\jd\rangle>0.5$ and $\langle\jh\cdot\jd\rangle>0.5$). The more recently the latest merger events occur, the stronger the gas-DM spin alignments 
become, which implies that the merger events have an effect of aligning the baryon spin axes of the galaxies toward the DM counterparts.  
Noting also that the $\jl$-$\jd$ and $\jh$-$\jd$ alignments exhibit little difference in strength between each other, we suspect that the cooling process in the cold gas 
should have little effect of weakening the gas-DM spin alignments.  
Figure \ref{fig:mdis1} and \ref{fig:del1} show how well the four $\am$-selected samples are controlled to yield almost identical mean values of $\log\tilde{M}$ 
and $\log(1+\delta)$ for three differences cases of the particle number cuts. It is now guaranteed that the different strengths of the baryon-DM spin 
alignments among the four $\am$-selected samples,  as witnessed in Figure \ref{fig:jjd}, are caused not by any difference in the galaxy total masses and environmental 
densities.

\subsection{Spin-shear alignments} 

Figure \ref{fig:jev1} shows how the spin orientations of the galaxy stellar, cold and hot gas components relative to the major principal directions of the 
local tidal tensors differ among the four $\am$-selected samples (top, middle and bottom panels, respectively).  A significant {\it alignment} signal is found only from the 
stellar spin axes of the galaxies whose latest merger events occur at $\am< 0.4$. The earlier the latest merger events occur, the stronger the $\js$-$\em$ alignments 
become, provided that $\am<0.4$. 
For the case of the galaxies that experience more recent merger events at $\am\ge 0.4$, however, their stellar spin axes seem to be randomly oriented with 
respect to the major principal directions of the tidal field. 

This result implies that the $\js$-$\em$ alignments should be generated by some physical process during the quiescent evolution period, 
and that the occurrence of the merger events plays the role of destroying the $\js$-$\em$ alignments. 
A comparison with the results shown in Figure \ref{fig:jjd} leads us to think that the $\js$-$\jd$ anti-alignments and the $\js$-$\em$ alignments must be generated 
by the same physical process, as envisaged by \citet{lee-etal22}. 
In contrast, the non-stellar gas spin axes of the galaxies yield only anti-alignment tendencies with the major principal directions of the tidal field, regardless of $\am$.
For the case of the cold gas, the anti-alignment  strength increases with $\am$, while for the case of the hot gas, no significant variation with $\am$ is found. 

The top panel of Figure \ref{fig:jdev} shows how the DM spin orientations relative to the major principal directions of the tidal tensors differ among the four $\am$-selected 
samples (top, middle and bottom panels, respectively). As can be seen in the top panel, the $\jd$-$\em$ anti-alignment does not exhibit a strong variation with the latest 
merger epochs for the case of $\am<0.66$, but a substantially weaker signal is found for the case of $\am\ge 0.66$, similar to the $\jh$-$\em$ anti-alignment 
(bottom panel of Figure \ref{fig:jev1}). 
A comparison of this result with those shown in Figure \ref{fig:jev1} casts two crucial implications. First, the merger events have only momentary weak effect of undermining the 
$\jh$-$\em$ and $\jd$-$\em$ anti-alignment tendencies. Second,  the hot gas and DM spin orientations should be insensitive to the physical process responsible for 
the generation of the $\js$-$\em$ alignments during the quiescent evolution period, while the merger events have an indirect effect of weakening the $\js$-$\em$ 
alignments and $\jl$-$\em$ anti-alignments by ending the quiescent period. 

Recall that the DM spin axes exhibit conspicuous mass-dependent transition phenomena: In the high-mass section, the DM spin orientations are preferentially aligned 
with the intermediate principal axes of the tidal fields while in the low-mass section they tend to be aligned with the minor principal directions \citep{lee-etal21}. 
Taking into account the occurrence of the DM spin transition phenomena, we divide the galaxies into the high-mass and low-mass groups containing those galaxies 
with $n_{d}\ge 10^{4}$ and $300\le n_{d}<10^{4}$, respectively.  Then, we investigate separately how the DM spin orientations of the high-mass and low-mass galaxies 
relative to the intermediate and minor principal axes vary with the latest merger epochs, the results of which are plotted in the middle and bottom panels of Figure 
\ref{fig:jdev}, respectively.  As can be seen, the strengths of the $\jd$-$\ei$ and $\jd$-$\en$ alignments do not show any significant variation with the latest merger epochs, 
which implies that the merger events are not responsible for the mass-dependent transition of the DM spin vectors relative to the principal directions of the tidal field. 
Figures \ref{fig:mdis2}-\ref{fig:del2} plots the same as Figures \ref{fig:mdis1}-\ref{fig:del1} but for four different cases of the particle number cuts, proving that the four controlled 
samples have no differences in the total mass and environmental density distributions, and thus that the detected trend of the spin-shear alignments is truly caused by the 
differences in the latest merger epochs. 

\subsection{Spin-shape and shape-shear alignments}

Figure \ref{fig:ju} shows how the strengths of the spin-shape alignments of the galaxy stellar, cold and hot gas, and DM components depend on the latest merger epochs 
(top, second from the top, second from the bottom, bottom panels, respectively). As can be seen in the top panel, the stellar components yield the strongest spin-shape 
alignments whose strength monotonically decreases with $\am$, The earlier the latest merger events occur, the stronger the $\js$-$\us$ alignments become. Note that the 
galaxies whose latest merger epochs occur at $\am \le 0.25$ yield $\langle\vert\js\cdot\us\vert\rangle\sim 1$, signaling the existence of almost perfect stellar spin-shape 
alignments. The physical process responsible for the peculiar $\js$-$\em$ alignments in the quiescent period must generate this almost perfect stellar spin-shear 
alignments, while the mergers ruin this strong alignments by interrupting the quiescent period. 

As shown in the bottom panel of Figure \ref{fig:ju}, the variation of the DM spin-shape alignment with $\am$ exhibits a directly opposite trend, increasing as the latest merger 
events occur more recently.  In other words, the mergers drive the DM components to have shapes whose minor axes become more strongly aligned with their spin vectors. 
A comparison with the result shown in the top panel of Figure \ref{fig:ju} leads us to think that the spin-shape alignments can be enhanced by two different mechanisms, 
the mergers and the physical process responsible for the $\js$-$\em$ alignments. The DM components are insensitive to the latter mechanism, while the stellar components 
are sensitive to both of the mechanisms.  Although the former mechanism has a direct effect of enhancing the spin-shape alignment even of stellar component, 
it also produces an indirect opposite effect by ending the quiescent period, only during which the latter mechanism effectively enhances the stellar spin-shape alignments.
This scenario explains why the stellar spin-shape alignments decrease with $\am$, while the DM counterparts exhibit an opposite trend. 

Meanwhile, the spin-shape alignments of the cold and hot gas components yield non-monotonic variations with the latest merger epochs, as shown in the second and third 
panels from the bottom of Figure \ref{fig:ju}. This result indicates that the non-stellar gas components are affected by both of the mechanisms and that the 
ranges of $\am$ at which the weakest spin-shape alignments are found correspond to the epochs when the two mechanisms counter-balance each other. 
Note that the hot gas components seem to be more vulnerable to the effect of the mergers while on the cold gas counterparts is more dominant the effect of 
the physical process responsible for the $\js$-$\em$ alignments. 

Figure \ref{fig:uev1} shows how the shape-shear alignments of the stellar, cold and hot gas, and DM components depend on the latest merger epochs 
(top, second from the top, second from the bottom, bottom panels, respectively). As can be seen, the shape vectors of the gas and DM components tend to 
align with the major principal axes of the tidal tensors to which their spin vectors tend to be perpendicular. This result implies that the shape axes of the 
gas and DM components are not good tracers of their spin directions. The shape-shear alignments of all of the four components decrease almost monotonically 
with $\am$, which indicates that the mergers have an effect of deviating the galaxy shape directions from the major principal axes of the local tidal tensors, 
regardless of the components. It is the DM components that exhibit the strongest shape-shear alignments while the weakest alignments are found 
from the cold gas components. 

Witnessing that the signals of the shape-shear alignments are significant in the entire range of $\am$ for the cases of the hot gas and DM, while 
the cold gas and stellar components yield no significant signals at $\am > 0.65$,  
we suspect that the cooling process of baryon particles occur most efficiently along the directions of maximum matter compression, which in consequence deviate the 
shape vectors of the cold gas and stellar components (i.e., the minor axes of the inertia momentum tensors) from the major principal axes of the local tidal tensors. 
Note also that the strengths and $\am$-dependence of the shape-shear alignments of the galaxy stellar components are very similar to those 
of their spin-shear alignments, as naturally expected from our finding of the very strong $\js$-$\us$ alignments (see Figure \ref{fig:ju}). 

\subsection{Triaxiality and sphericity}

Figures \ref{fig:tri} and \ref{fig:sph} show how the merger events affect the morphological shapes of the galaxy stellar, cold and hot gas, and DM components 
(top, second from the top, second from the bottom, and bottom panels, respectively). 
As can be seen,  the galaxy stellar components have conspicuously oblate shapes ($\tris< 0.5$), unlike the other three components all of which exhibit prolate shapes 
($\tril,\ \trih,\ \trid> 0.5$). The earlier the latest merger events occur, the more oblate shapes the galaxy stellar components have, indicating that the galaxy stellar 
components develop oblate shapes during the quiescent period, but evolve into less oblate and more elongated shapes through the mergers.  
The mergers also reorganize the galaxy cold gas and DM distributions of the galaxies to have more prolate and aspherical shapes, while the hot gas component of the galaxies 
seem to be least susceptible to the merger events, with the mean values of $\trih$ and $\sphh$ varying only weakly with $\am$.  

Among the four components, it is the cold gas components that have the most prolate and most aspherical shapes, while it is the stellar components that have the most oblate shapes. 
Note also that the cold gas components are more aspherical than the stellar counterparts, with $\sphl$ decreasing almost monotonically with $\am$.  Given this palpable 
difference between the morphologies of the stellar and cold gas components, we suspect that the cooling process in the cold gas should not directly contribute to the generations 
of the peculiar $\js$-$\em$ alignment and the strongest $\js$-$\us$ alignments. In other words, the physical process through which the gas particles cool down and evolve into stars is 
unlikely to be the main mechanism that drives the stellar spin and shape vectors to align with the directions of maximum matter compression. If the gas cooling process were to 
be the main mechanism, the morphologies of the cold gas components would be similar to those of the stellar counterparts, having less prolate shapes than the hot gas and DM 
components. 

\section{Summary and discussion}\label{sec:con}

\begin{figure}[ht]
\begin{center}
\plotone{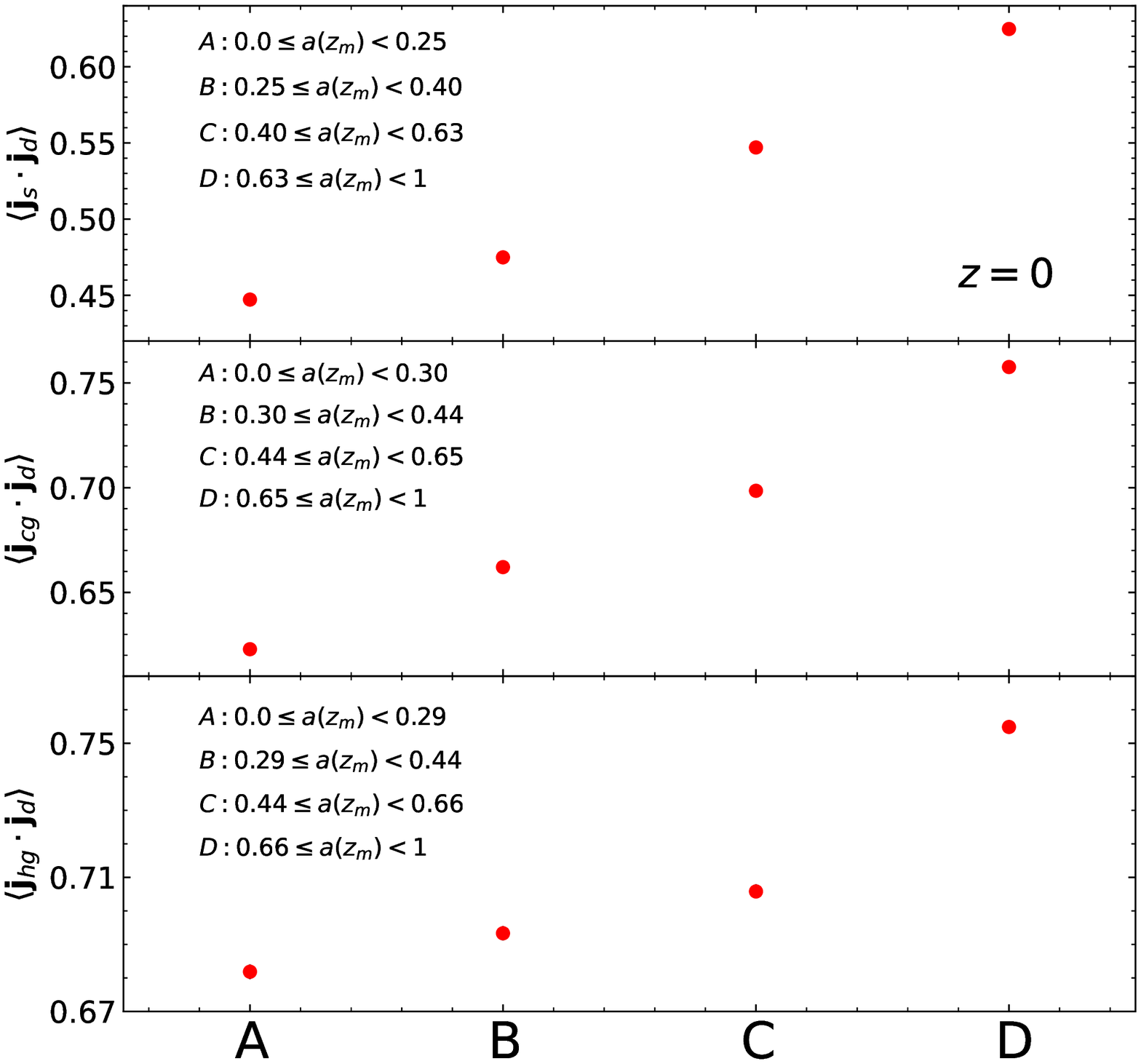}
\caption{Mean values of the cosines of the angles of the galaxy stellar, cold gas and hot gas spin axes
from their DM spin axes (top, middle and bottom panels, respectively), at $z=0$, averaged over  each of the four 
controlled samples classified by the latest merger epochs, $\am$. The errorbars are not visible due to their tiny sizes. 
The earlier the galaxies experience their latest mergers, the weaker the spin alignments between their baryon 
and DM components become.}
\label{fig:jjd}
\end{center}
\end{figure}

\begin{figure}[ht]
\begin{center}
\plotone{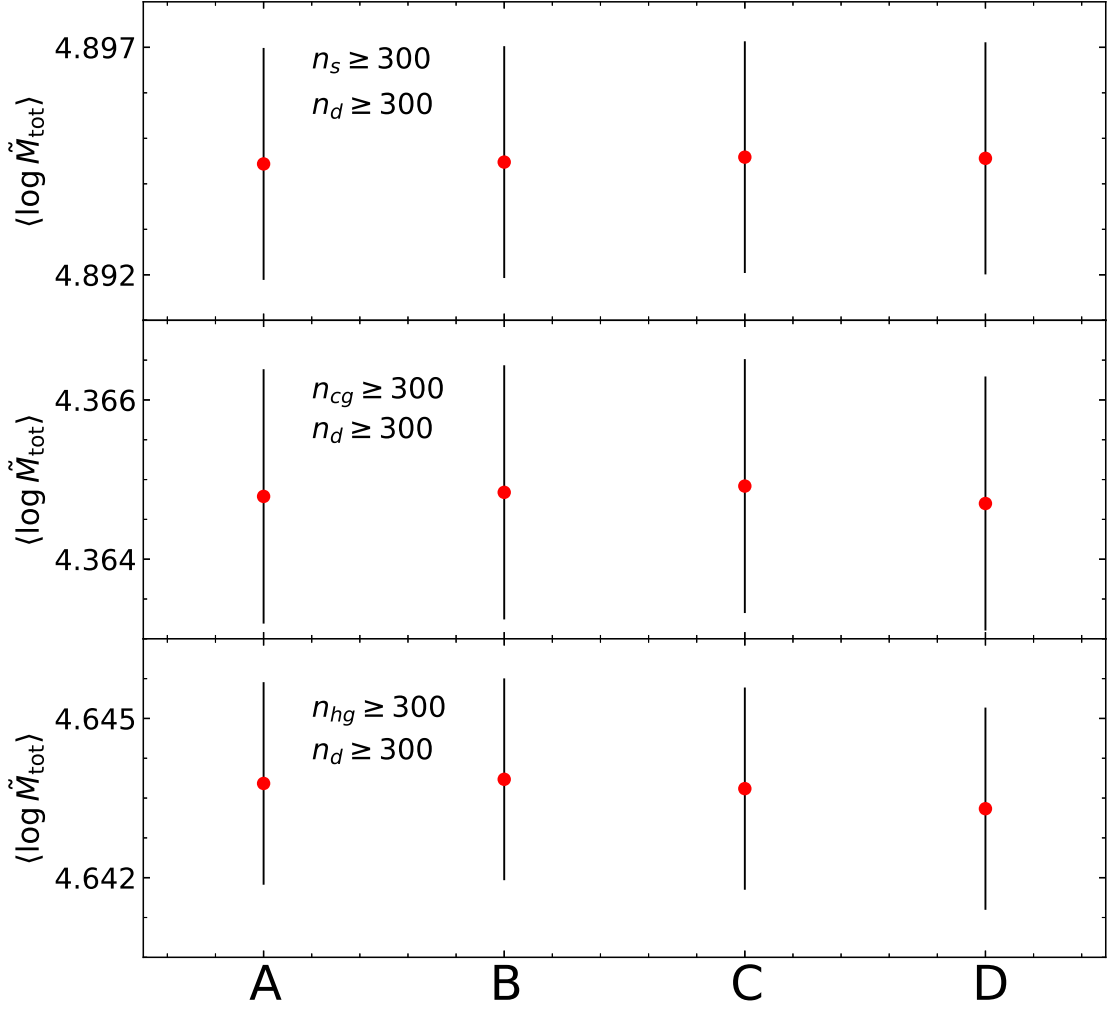}
\caption{Mean values of the galaxy total masses averaged over each of the four controlled samples for three different cases 
of the particle number cuts: $n_{s}, n_{d}\ge 300$ (top panel); $n_{cg}, n_{d}\ge 300$ (middle panel); $n_{hg}, n_{d}\ge 300$ 
(bottom panel).}
\label{fig:mdis1}
\end{center}
\end{figure}

\begin{figure}[ht]
\begin{center}
\plotone{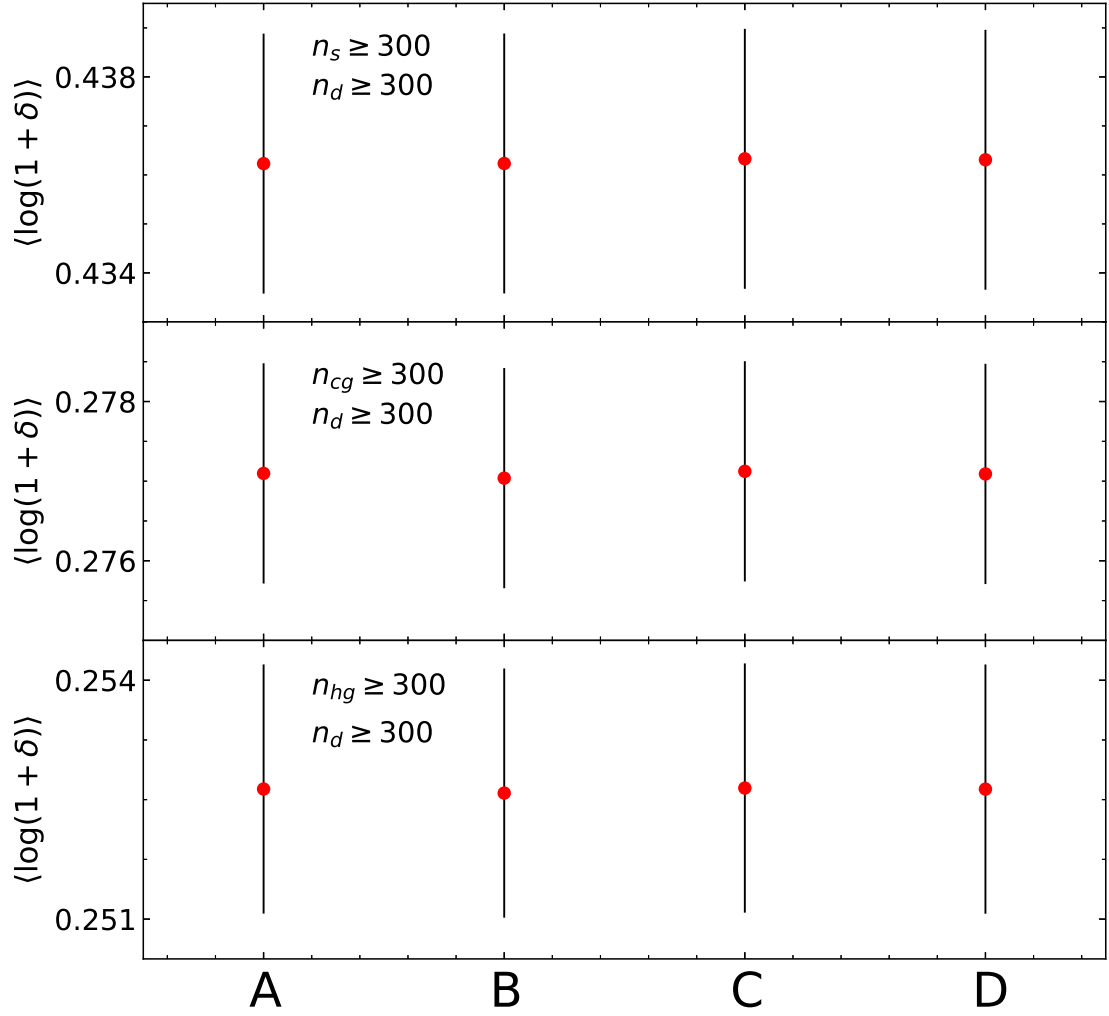}
\caption{Mean values of the environmental density contrasts averaged over each of the four controlled samples for the 
three different cases of the particle number cuts.}
\label{fig:del1}
\end{center}
\end{figure}

\begin{figure}[ht]
\begin{center}
\plotone{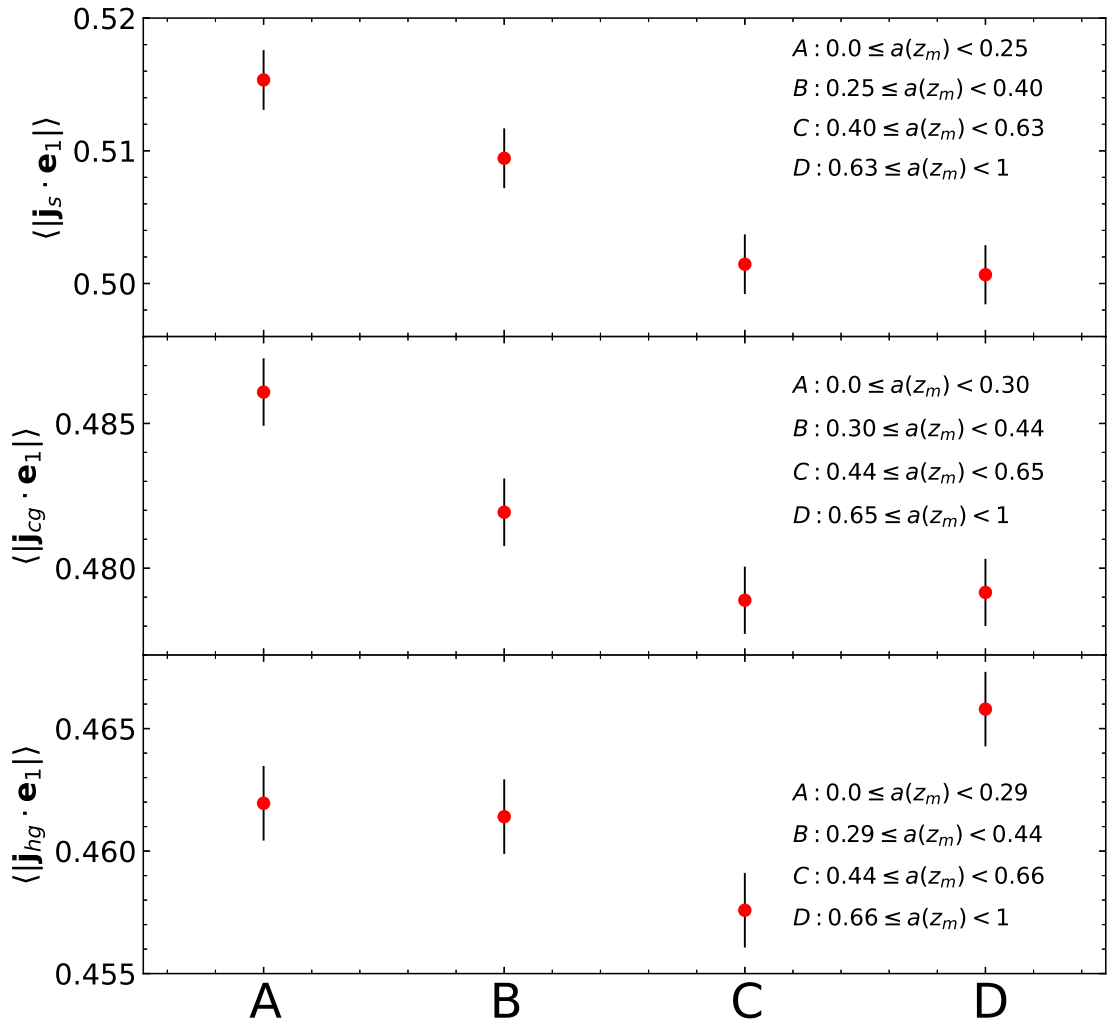}
\caption{Mean absolute values of the cosines of the angles of the galaxy stellar, cold gas, and hot gas spin vectors of the galaxies 
from the major principal directions of the local tidal tensors (top, middle, and bottom panels, respectively) 
averaged over each of the four controlled $a(z_{m})$-selected samples. The earlier the galaxies experience their latest mergers, 
the stronger the peculiar tidal connections of their stellar components become.}
\label{fig:jev1}
\end{center}
\end{figure}

\begin{figure}[ht]
\begin{center}
\plotone{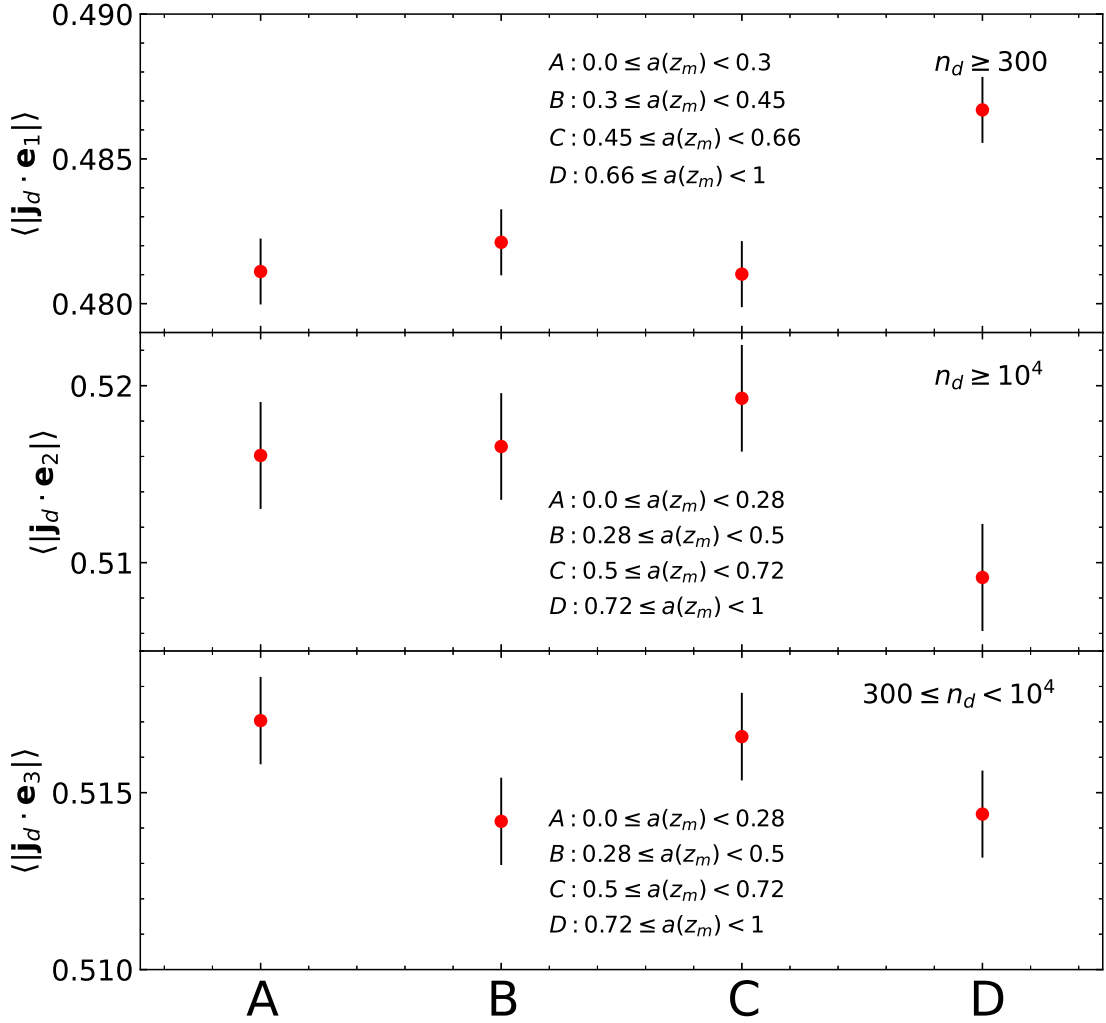}
\caption{Mean absolute values of the cosines of the angles of the galaxy DM spin vectors from the major, intermediate and 
minor principal directions of the local tidal tensors (top, middle and bottom panels, respectively) averaged over each of the four 
controlled $a(z_{m})$-selected samples. No strong dependence of the alignment strength between the galaxy DM spins 
and the principal axes of the local tidal fields on the latest merger epochs.}
\label{fig:jdev}
\end{center}
\end{figure}

\begin{figure}[ht]
\begin{center}
\plotone{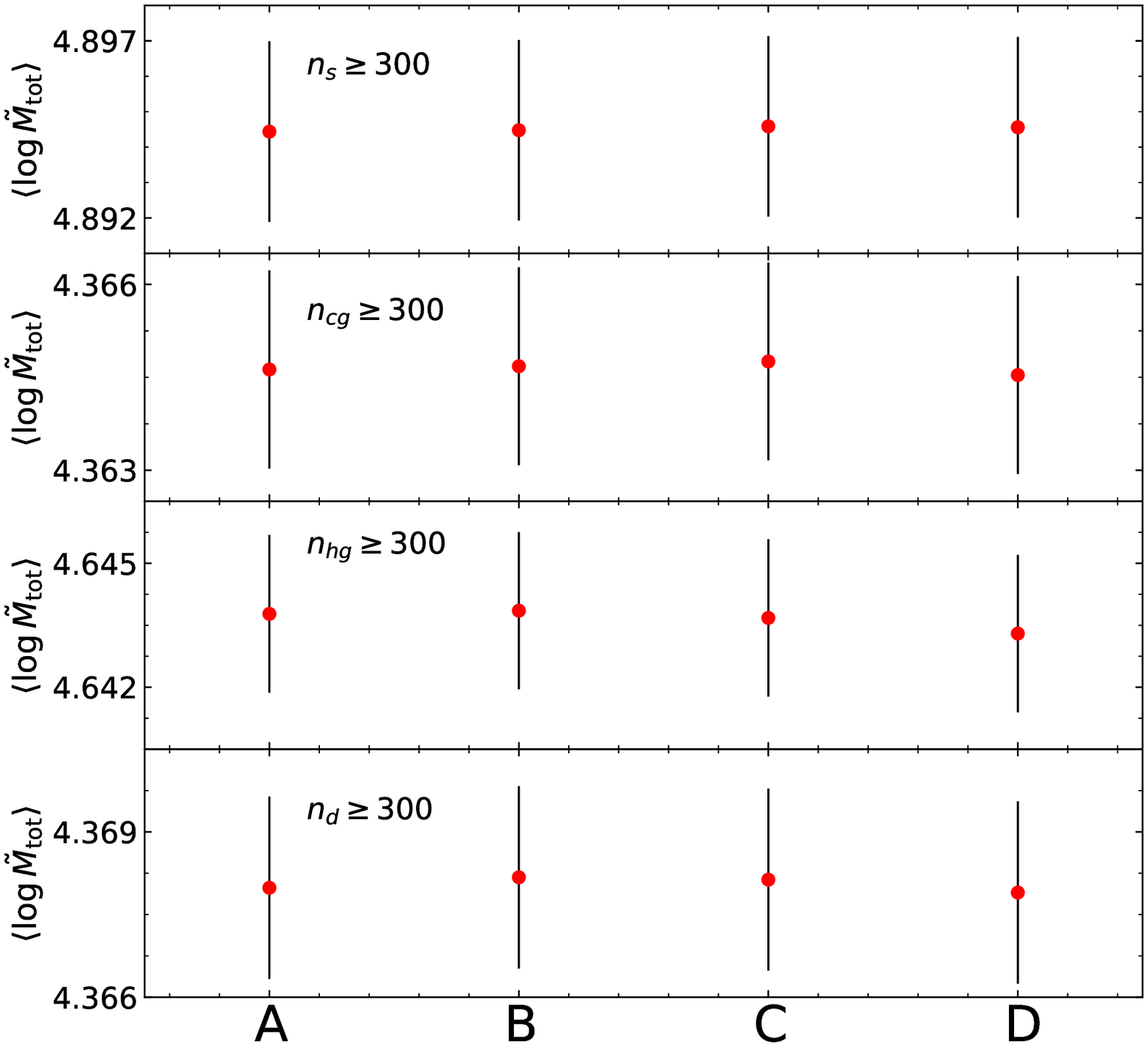}
\caption{Same as Figure \ref{fig:mdis1} but for four different cases of the particle number cuts: 
$n_{s}\ge 300$ (top panel); $n_{cg}\ge 300$ (second panel from the top); $n_{hg}\ge 300$ (third panel from the top); 
$n_{d}\ge 300$ (bottom panel).}
\label{fig:mdis2}
\end{center}
\end{figure}

\begin{figure}[ht]
\begin{center}
\plotone{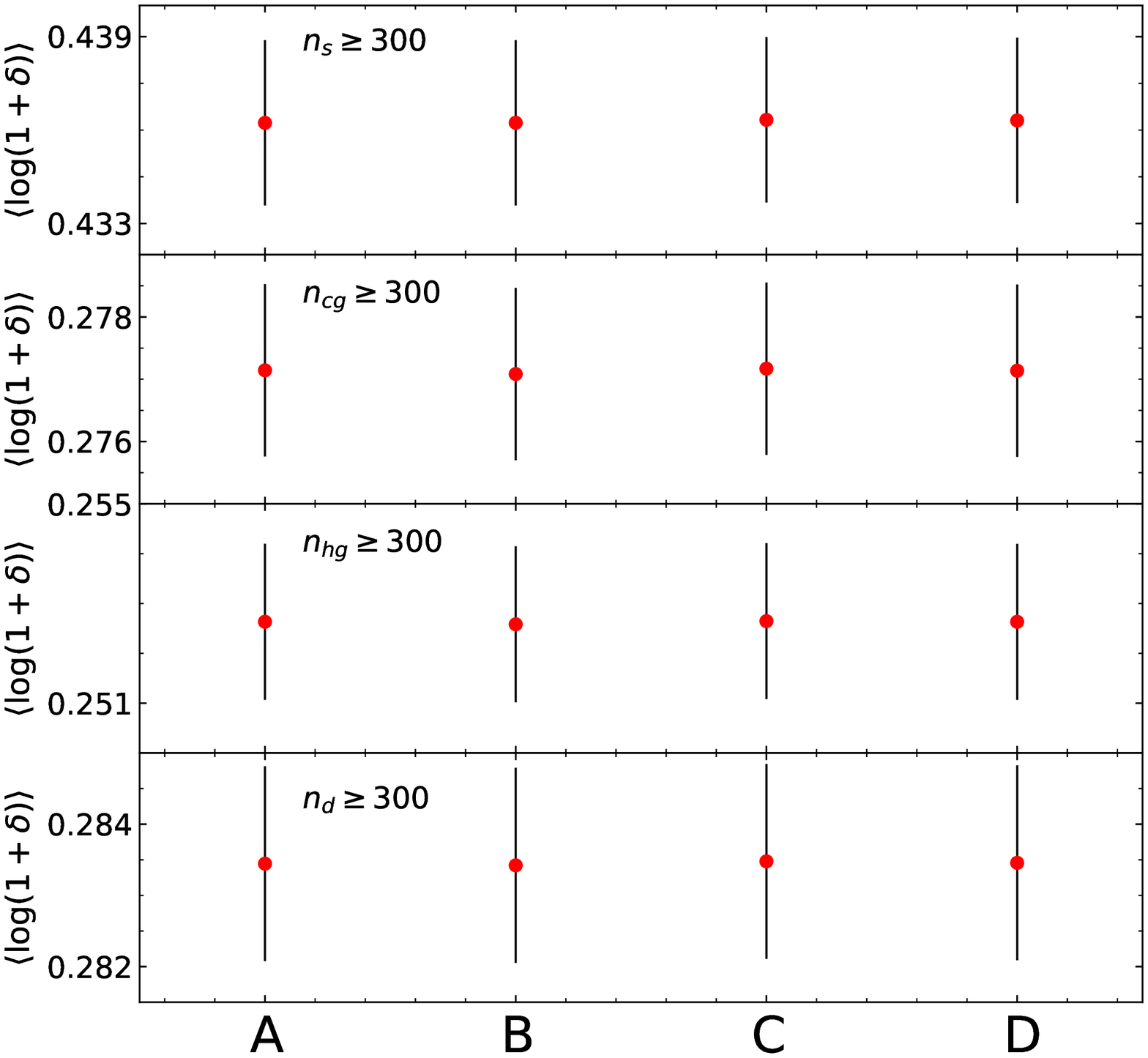}
\caption{Same as Figure \ref{fig:del1} but for four different cases of the particle number cuts: 
$n_{s}\ge 300$ (top panel); $n_{cg}\ge 300$ (second panel from the top); $n_{hg}\ge 300$ (third panel from the top); 
$n_{d}\ge 300$ (bottom panel).}
\label{fig:del2}
\end{center}
\end{figure}

\begin{figure}[ht]
\begin{center}
\plotone{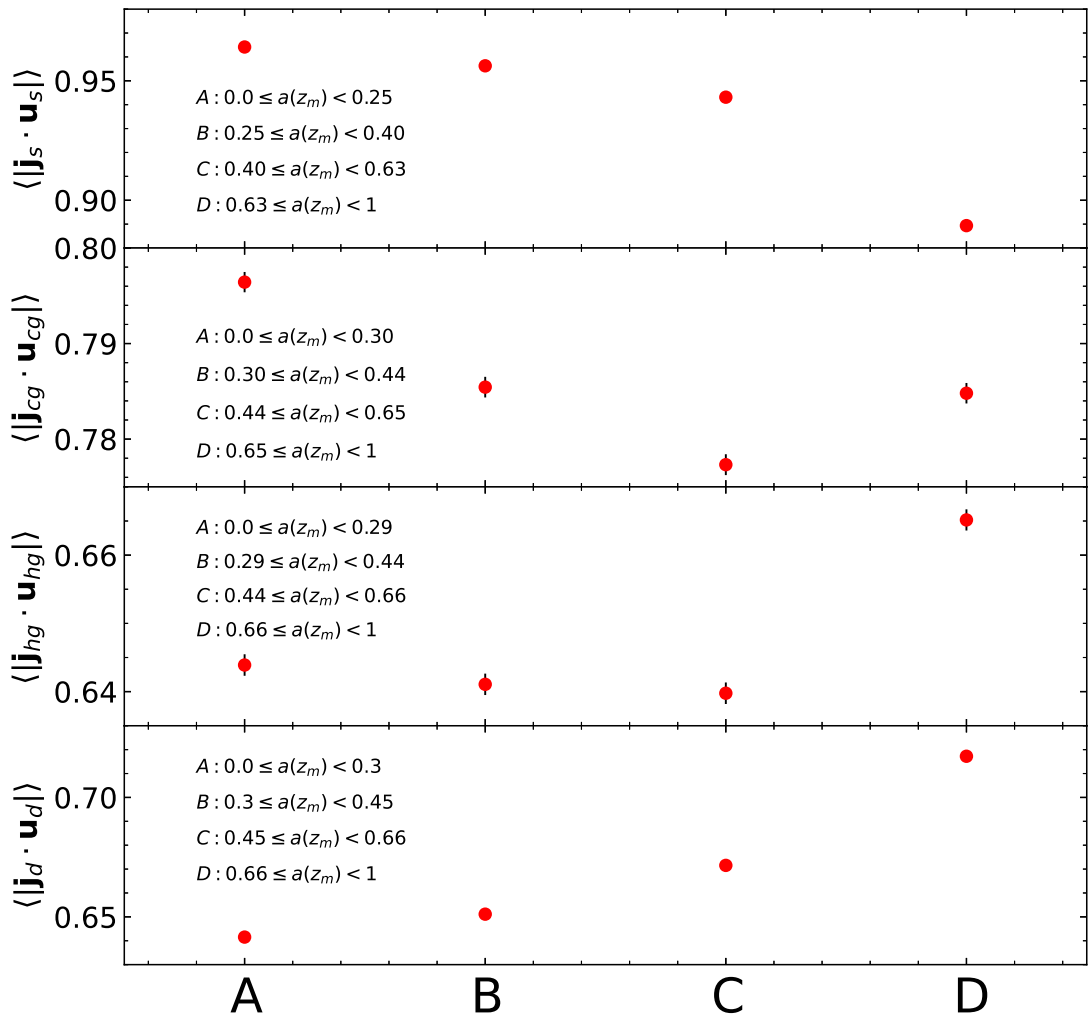}
\caption{Mean values of the dot products between the spin and shape vectors of the galaxy stellar, cold gas, hot gas 
and DM components (top, second from the top, third from the top and bottom panels, respectively), averaged over 
each of the four controlled $a(z_{m})$-selected samples. In the top and bottom panels, the errorbars are invisible due to 
their tiny sizes. The earlier the latest mergers occur, the stronger (the weaker) the alignments between the galaxy stellar (DM) 
spin and shape vectors become.}
\label{fig:ju}
\end{center}
\end{figure}

\begin{figure}[ht]
\begin{center}
\plotone{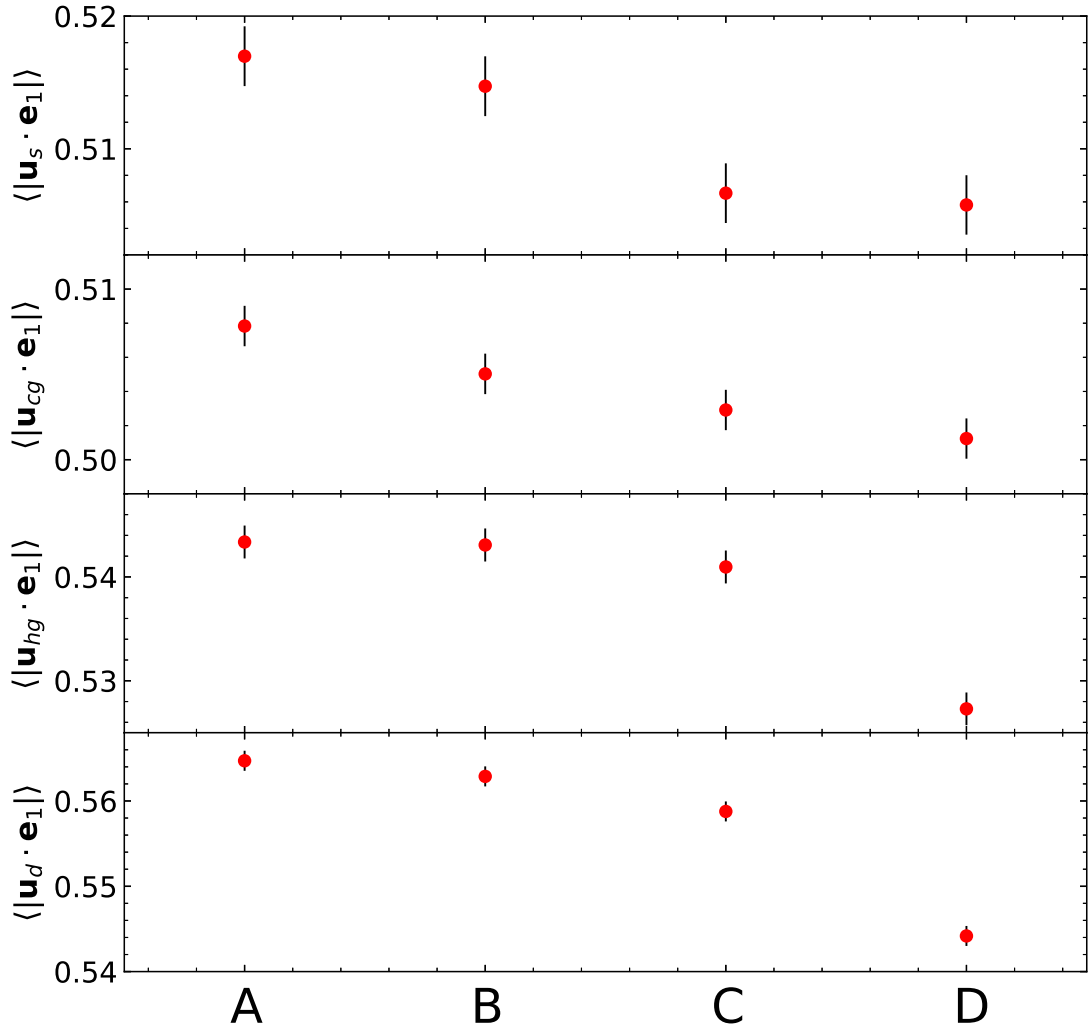}
\caption{Mean absolute values of the dot products of the galaxy stellar, cold gas, hot gas, DM shape vectors 
from the major principal directions of the local tidal tensors  (top, second from the top, third from the top and bottom 
panels, respectively), averaged over each of the four controlled $a(z_{m})$-selected samples.
The earlier the latest mergers occur, the stronger alignments the galaxies exhibit between their spin and shape vectors.}
\label{fig:uev1}
\end{center}
\end{figure}

\begin{figure}[ht]
\begin{center}
\plotone{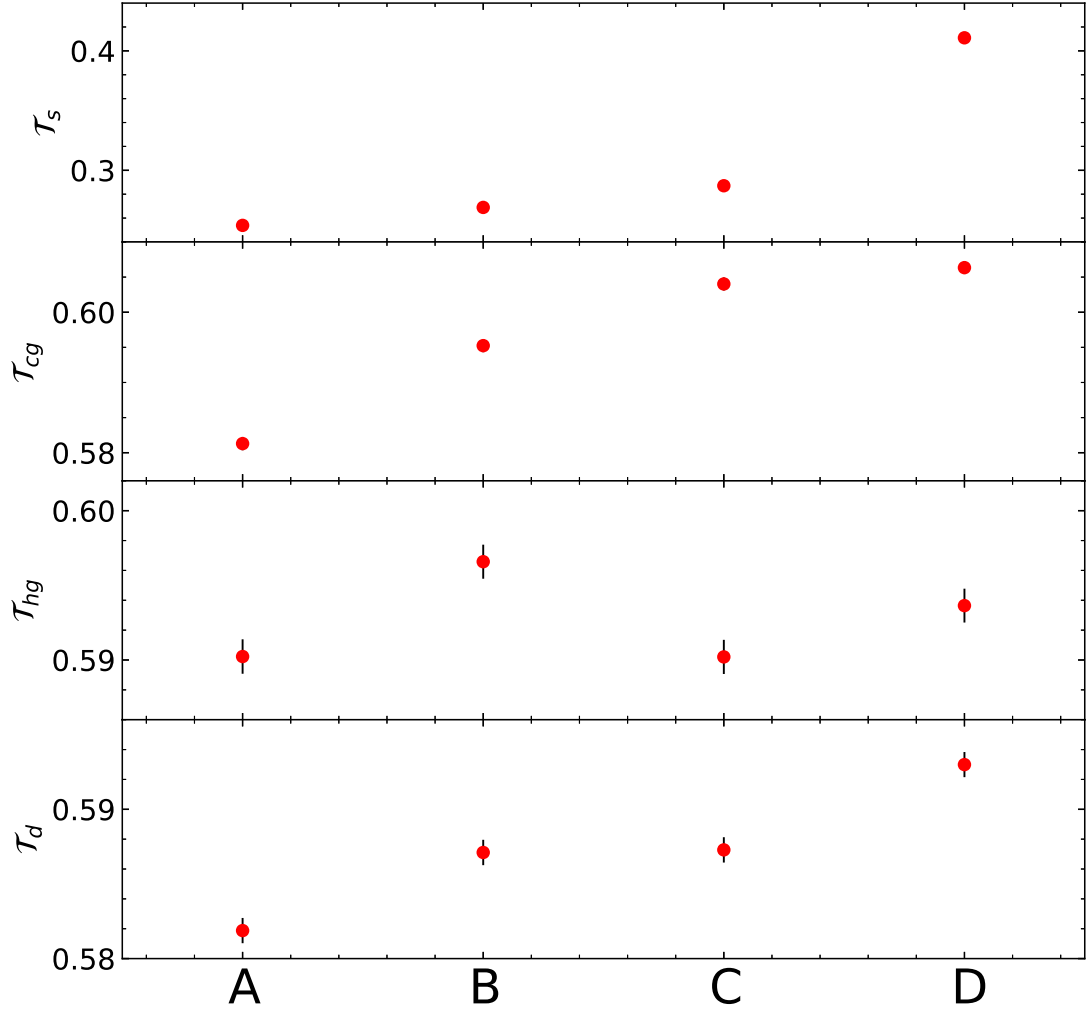}
\caption{Mean values of the triaxiality of the galaxy stellar, cold gas, hot gas and DM components 
averaged over the four controlled $a(z_{m})$-selected samples. The errorbars in the top two panels are 
invisible due to their tiny sizes. The earlier the latest mergers occur, the less triaxial shapes the galaxies have.}
\label{fig:tri}
\end{center}
\end{figure}

\begin{figure}[ht]
\begin{center}
\plotone{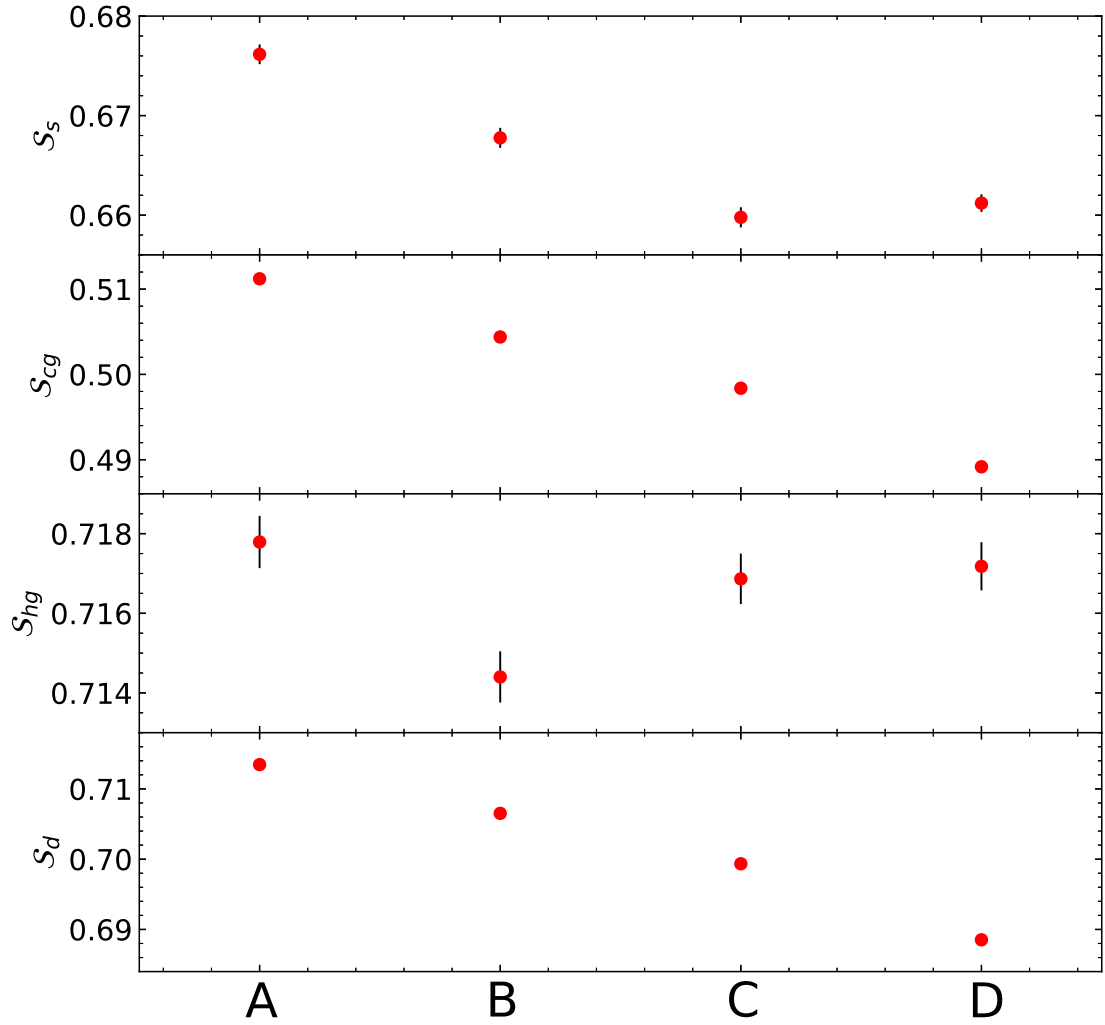}
\caption{Same as Figure \ref{fig:tri} but of the sphericity. The errorbars in the second panel from the top are 
invisible due to their tiny sizes. The earlier the latest mergers occur, the more spherical shapes the galaxies have.}
\label{fig:sph}
\end{center}
\end{figure}

Analyzing separately the stellar, cold and hot gas, and DM components of the well-resolved galaxies from the TNG 300-1 simulation of the IllustrisTNG project, 
we have explored how the strengths and tendencies of the spin-shear, spin-shape, and shape-shear alignments of each component depend on the latest merger epochs at 
$z=0$.  The effects of the environmental density and total mass differences on the strengths of these alignments have been properly eliminated by controlling the galaxy 
samples. The key results from this exploration and their implications are summarized in the following.
\begin{itemize}
\item
The peculiar alignments of the galaxy stellar spins with the major principal axes of the tidal field and their anti-alignments with the DM counterparts can be developed only 
during the quiescent evolution period when the galaxies experience no recent mergers (Figures \ref{fig:jjd}-\ref{fig:jev1}).  The merger events have an effect of 
destroying the peculiar stellar spin alignments by orienting the stellar spin axes toward the DM counterparts that are preferentially aligned with the directions perpendicular 
to the major principal axes of the local tidal tensors. The physical process responsible for the generation of this peculiar stellar spin alignments must affect only the stellar 
and cold gas components but not the hot gas and DM components. 
\item
No strong variation of the strengths with the latest merger epoch is exhibited by the DM spin alignments of high-mass and low-mass galaxies with the intermediate and 
minor principal directions of the local tidal tensors, respectively (Figures \ref{fig:jdev}-\ref{fig:del2}), which implies that the merger events are not the primary mechanism 
for the generation of the spin transition phenomenon of galactic halos \citep[c.f.,][]{pic-etal11,cod-etal12,dub-etal14,kro-etal19}.  
\item 
During the quiescent period, the galaxy stellar components develop oblate shapes whose minor axes are almost perfectly aligned with their spin directions 
(Figures \ref{fig:ju}-\ref{fig:tri}). The stellar shape-shear alignments mirror very well the stellar spin-shear alignments in strengths and behaviors as a function 
of the latest merger epoch (Figure \ref{fig:uev1}). The physical process responsible for the peculiar stellar spin alignments should be closely linked with the 
almost perfect stellar spin-shape alignments. 
\item
The non-stellar gas and DM components have prolate shapes whose minor axes are not so perfectly aligned with their spin directions as the stellar components. 
The merger events enhance the DM spin-shape alignments, while weakening the stellar counterparts (Figure \ref{fig:ju}). This striking difference between the stellar and 
DM components is likely caused by the dual roles of a merger event. Although it has the effect of enhancing the spin-shape alignments of each component, 
the occurrence of a merger event ends the quiescent period at the same time, which incapacitates the same enhancing effect of the physical process on the 
stellar spin-shape alignments. The duality of the merger effects is well reflected by the non-monotonic variation of the cold gas spin-shape alignments with the latest 
merger epochs. 
\end{itemize}

Our results provides a counter evidence against the prevalent scenario that the spin transition of galactic halos should be ascribed to the merger effects 
\citep{ara-etal07,hah-etal07,pic-etal11,cod-etal12,tro-etal13,dub-etal14,cod-etal18,gan-etal18,kro-etal19}. According to this scenario, the spin axes of high-mass galaxies 
acquire a tendency of being aligned with the directions perpendicular to the host filaments, as they evolve through anisotropic mergers along the filaments. 
Whereas the low-mass galaxies have their spin directions aligned with the directions parallel to the filaments, as they undergo not so frequent mergers as the high-mass 
counterparts. If this scenario were true, then the strengths of the spin alignments of the high-mass galaxies with the intermediate principal directions of the 
tidal tensors (perpendicular to the filament axes) would yield a signal of significant variation with the latest merger epochs. However, no such signal has been detected 
by our analysis, which indicates that the mergers contribute very little to establishing the spin transition of the galactic halos. 

Rather, our finding is consistent with the tidal torque picture that the DM spin alignments of high-mass galaxies with the filament axes originate from the tidal interactions 
at the proto-galactic stages \citep{LP00,LP01,jon-etal10,mot-etal21,lee-etal21,lee-etal22}.  
In the subsequent evolution, the occurrence of mergers could significantly change the DM spin directions of galactic halos without destroying 
the tidally induced alignments, since the DM spin angular momenta of the merged halos are in fact the transferred orbital angular momenta that are also aligned with 
the intermediate principal axes of the tidal field on larger scales. In other words, the merger events change the scale of the tidal field whose principal axes the DM spin 
directions are aligned with.  Regarding the DM spin alignments of low-mass galaxies with the minor principal axes, the tidal torque theory provides no physical  
answer to the critical question of what mechanism produces it. However, given our result that their strengths also exhibit no variation with the latest merger 
epochs, their origin should not be the occurrence of the merger events, either, as suspected by \citet{lee-etal20}. 
 
Our result also challenges the conventional assumption that the minor axes of the galaxy shapes is a good proxy of their spin directions. Under this assumption, the observed 
stellar shape alignments of the galaxies with the cosmic web were often regarded and interpreted in the previous works as the evidences for the existence of the DM spin 
alignments of galactic halos \citep[e.g.,][]{jon-etal10,lee-etal18,kro-etal19,lee19,wel-etal20,kra-etal21,mot-etal21}, since the former is more readily observable while the latter is 
what a theory or a numerical experiment can model and predict. 
As revealed by our work, however, only for the case of the galaxy stellar components, their shape-shear alignments mirror well the spin-shear counterparts, while for the 
case of the non-stellar gas and DM components of the galaxies, the former notably differs from the latter, especially in variation with the latest merger epochs. 
In other words, the theoretical model for the DM spin-shear alignments of galactic halos should not be directly compared with the observed stellar shape-shear alignments, 
since the two phenomena are likely to originate from different mechanisms. 

Yet, it is worth mentioning that our results on the peculiar tidal connections are contingent upon our choice of $2R_{1/2}$ as the radial distance cut-off for the determinations 
of the galaxy stellar spin directions. As previously shown in \cite{chi-etal16}, the strength and tendency of the stellar spin and shape alignments of the galaxies sensitively 
vary with the choice of the cut-off or weights on the radial distances of the stellar components \citep[see, also][]{jag-etal22}.
The discrepancy of our result with the previous works which found quite similar tendency between the stellar and DM spin axes of the galaxies in their alignments with 
the cosmic web \citep{cod-etal18} may be caused by the difference in the way that the stellar angular momentum vectors were measured. 
Unlike the previous work where the subhalos having $50$ or more stellar components without putting any radial distance cutoff,  the current analysis has 
taken a more conservative approach in line with the observational analyses, including only those subhalos having $300$ or more stellar components within $2R_{1/2}$. 

Nevertheless, the existence of the peculiar tidal connection of the stellar spins of the galaxies in quiescent evolution and their almost perfect spin-shape alignments hints 
that it might be even more plausible to reconstruct the local tidal and density fields from the observable stellar shapes of the galaxies with an algorithm 
similar to the one developed by \citet{LP00} and recently used in practice by \citet{mot-etal21}. In these previous works, the weak alignments between the DM 
spin axes of the galaxies and the intermediate principal axes of the tidal fields is assumed to reflect well the anisotropic tidal fields.  This algorithm, however, 
has a limited usefulness due to several practical difficulties: First, the DM spin axes are not readily observable unlike the stellar counterparts. 
Second, the alignments between the DM spins and the intermediate principal axes of the tidal fields can be found only from the high-mass galaxies but not from 
the low-mass galaxies due to the spin transition phenomena. Third, in the sheet and filamentary environments, the intermediate principal axes of the tidal fields are 
degenerated with the other principal axes. 

If this algorithm could be modified to accommodate the peculiar stellar spin alignments of the galaxies with the major principal 
axes of the tidal fields and to use it for the reconstruction of the tidal and density fields, then the aforementioned practical difficulties would be avoided.  
The stellar shape alignments of the galaxies are readily observable, the peculiar stellar spin alignments do not exhibit transition phenomena on the galactic scale, 
and the major principal axes of the tidal fields do not suffer from the degeneracy even in the filamentary or sheet environments. 
Our future work will be in the direction of developing such an algorithm.

In conclusion, we have found several key clues to the physical process that generates the peculiar alignments of the galaxy stellar spins with the major principal 
axes of the local tidal field. We believe that these clues will provide deeper insights to understand and model the physical process, which must precede any attempt to 
connect the observable shape-shear alignments of the galaxy stellar components to the initial conditions of the universe \citep[e.g.,][]{LP00,TO20,mot-etal21}.

\acknowledgments

We thank an anonymous referee whose thoughtful review helped us improve the original manuscript. 
The IllustrisTNG simulations were undertaken with compute time awarded by the Gauss Centre for Supercomputing (GCS) 
under GCS Large-Scale Projects GCS-ILLU and GCS-DWAR on the GCS share of the supercomputer Hazel Hen at the High 
Performance Computing Center Stuttgart (HLRS), as well as on the machines of the Max Planck Computing and Data Facility 
(MPCDF) in Garching, Germany. 
JL acknowledges the support by Basic Science Research Program through the National Research Foundation (NRF) of Korea 
funded by the Ministry of Education (No.2019R1A2C1083855).

\clearpage
\clearpage
\clearpage
\clearpage
\clearpage
\clearpage
\clearpage
\clearpage
\clearpage
\clearpage
\clearpage
\end{document}